\documentclass[onecollarge,12pt]{style}
\smartqed  % flush right qed marks, e.g. at end of proof
\usepackage{graphicx}
\usepackage{mathptmx}
\usepackage{amssymb}
\usepackage{amsmath}
\usepackage{graphicx}% Include figure files
\usepackage{bm}
\journalname{Few-Body Systems}

\begin{document}

\title{The $^4$He trimer as an Efimov system$^{*\dagger}$\thanks{$^*$For a special issue
of \textit{Few-Body Systems} devoted to Efimov physics.}\thanks{$^\dagger$This work was
supported by the Deu\-t\-s\-che For\-s\-ch\-ungs\-ge\-mein\-schaft
(DFG), the Heisenberg-Landau Program, the \mbox{\hspace*{0.5em}Alexander} von Humboldt
Foundation, and the Russian Foundation for Basic Research.}}

\titlerunning{The $^4$He trimer as an Efimov system}     % if too long for running head

\author{E. A. Kolganova  \and A. K. Motovilov \and W. Sandhas}

\authorrunning{E.\,A.\,Kolganova, A.\,K.\,Motovilov, W.\,Sandhas} % if too long for running head

\institute{Elena A. Kolganova   \at
              Bogoliubov Laboratory of
              Theoretical Physics, JINR, Joliot-Curie 6, 141980 Dubna, Moscow Region, Russia\\
               \email{kea@theor.jinr.ru}
           \and
              Alexander K. Motovilov
              \at
              Bogoliubov Laboratory of
              Theoretical Physics, JINR, Joliot-Curie 6, 141980 Dubna, Moscow Region, Russia\\
              \email{motovilv@theor.jinr.ru}
           \and
           Werner Sandhas \at
              Physikalisches Institut, Universit\"at Bonn,
              Endenicher Allee
              11-13, D-53115 Bonn, Germany\\
              \email{sandhas@physik.uni-bonn.de}
}

\dedication{Dedicated to the 40th anniversary of the Efimov effect}

\date{Date: April 17, 2011}

\maketitle

\begin{abstract}
We review the results obtained in the last four decades which
demonstrate the Efimov nature of the $^4$He three-atomic system.
\end{abstract}

\section{Introduction}

For many years the existence of a bound state of two $^4$He atoms
was an open problem. Some potential models predicted  such a state
\cite{BruchMcGee1970,Aziz79,UangStwalley} while the others did not
\cite{DeBoer1958,Beck1968}. However practically all more or less
realistic helium-helium potentials generated a very large atom-atom
scattering length of about 90--100 {\AA} or even more (see, e.g.,
\cite[Table~I]{Aziz95}). As soon as the Efimov effect has been
discovered \cite{VEfimovYaF1970,VEfimov1970,VEfimov1973}, it was to
be expected that the system of three $^4$He atoms possesses bound
states of Efimov type. For the first time this idea has been
suggested and substantiated by Lim, Duffy, and Damert
\cite{LimDuffy1977}, just seven years after Efimov's first works on
his effect \cite{VEfimovYaF1970,VEfimov1970}. It is the Efimov
effect that distinguishes the $^4$He atoms from the atoms of all
other noble gases, and makes the $^4$He clusters especially
attractive objects of experimental and theoretical studies.

Almost all realistic He-He-potentials constructed in the 1970s and
later supported $^4$He$_2$ binding, although the binding energies
may differ by tens of times \cite{Aziz79,Anderson1993,Bishop1977}.
The semi-empirical Aziz \textit{et al.} potentials
\cite{Aziz87,Aziz91} are considered particularly adequate, as well
as the purely theoretical TTY potential by Tang, Toennies, and Yiu
\cite{TTY}. Compared to the others, the LM2M2 potential
\cite{Aziz91} seems to be most often used in $^4$He trimer
calculations of the last decade.  Besides these potentials, we also
mention the SAPT potentials developed by Korona \textit{et al.}
\cite{Korona}, by Janzen and Aziz \cite{Aziz97}, and by Jeziorska
\textit{et al.} \cite{Jeziorska2007}. All potentials
\cite{Aziz87,Aziz91,TTY,Korona,Aziz97,Jeziorska2007} support a
single bound state of two $^4$He atoms with a binding energy of
1.3--1.9 millikelvin (mK). For convenience, we collect in Table
\ref{tableDimerLen} the $^4$He$_2$ binding energies and
$^4$He--$^4$He scattering lengths obtained with the Aziz \textit{et
al.} potentials HFD-B \cite{Aziz87}, LM2M2 \cite{Aziz91}, and with
the TTY potential by Tang, Toennies, and Yiu \cite{TTY}. Similarly
to the SAPT potentials \cite{Korona,Aziz97,Jeziorska2007}, these
three potentials predict exactly two bound states for the $^4$He
trimer. The HFD-B potential gives about 133~mK for the ground state
energy \cite{MSSK,RoudnevYak,Barletta,BlumeGreene} and about 2.74~mK
for the energy of the excited state  \cite{MSSK,RoudnevYak}. The
corresponding results for the TTY potential are shown in Table
\ref{tableTrimers}. The respective energies obtained for the LM2M2
potential are practically the same as for the TTY potential (see
Table \ref{tableTrimers1}), that is, they are close to 126 mK for
the ground state and close to 2.28 mK for the
excited one~\cite{MSSK,RoudnevYak,Barletta,BlumeGreene,%
BlumeGreene113,Lazauskas,Nielsen,Bressanini,SalciLevin,KolganovaRC2010,%
Orlandini2009,Barletta09,Kolganova2010}. For the $^4$He trimer
binding energies obtained with the SAPT potentials we refer to the
calculations in \cite{Barletta,Barletta09,Roudnev2,SunoEsry2008}.

Results on the $^4$He atom -- $^4$He dimer scattering length and
phase shifts with realistic atom-atom potentials are less numerous.
In this respect we refer to
\cite{MSSK,BlumeGreene,Lazauskas,Barletta09,MSK-CPL,KMS-JPB,Roudnev1,PRA04,SunoEsry2008}.
For a discussion of problems of convergence arising in $^4$He --
$^4$He$_2$ scattering calculations see \cite[Section 5]{KMS-EChAYa}.
The $^4$He atom -- $^4$He dimer scattering lengths available for the
TTY and LM2M2 potentials are shown in Tables~\ref{tableTrimers}
and~\ref{tableTrimers1}, respectively.

By now, it is already rather well established that, if the  $^4$He
trimer excited state exists, then it should be of Efimov nature. As
already mentioned, the appearance of the Efimov effect in the $^4$He
three-atom system was conjectured in \cite{LimDuffy1977} where the
$^4$He$_3$ excited state binding energy has been calculated for the
first time by means of the Faddeev integral equations. Even more
convincing arguments in favor of this phenomenon were presented by
Cornelius and Gl\"ockle \cite{Gloeckle} who also employed the
momentum space Faddeev equations. Ten years later the conclusions of
\cite{Gloeckle} were strongly supported in \cite{EsryLinGreene} and
\cite{KMS-JPB}. The calculations of \cite{EsryLinGreene} were based
on the adiabatic hyperspherical expansion in three-body
configuration space, while the hard-core version of the
two-dimensional Faddeev differential equations has been used in
\cite{KMS-JPB}. References
\cite{Bedaque1999,Frederico1999,Yama2002,BraHam2003,Penkov2003,PenkovSandhas2006,Platter2006,Shepard2007}
suggest that the $^4$He$_3$ ground state itself may be considered as
an Efimov state since, given the $^4$He-$^4$He atom-atom scattering
length, both the $^4$He$_3$ ground-state and excited-state energies
lye on the same universal scaling curve (for details, see, e.g.,
\cite[Sections 6.7 and 6.8]{BraHam2006}).

Experimentally, $^4$He dimers have been observed for the first time
in 1993 by the Minnesota group~\cite{DimerExp}, and in 1994 by
Sch\"ollkopf and Toennies \cite{Science}. Along with the dimers, the
experiment \cite{Science} established also the existence of $^4$He
trimers. A first experimental estimate for the size of the
$^4$He$_2$ molecule has been given in \cite{DimerExp1}. According to
this reference, the root mean square distance between $^4$He nuclei
in the $^4$He dimer is equal to $62\pm10$\,{\AA}. Several years
later, the bond length for $^4$He$_2$ was measured again by Grisenti
{\em et al.} \cite{Grisenti-exp-2000} who found for this length the
value of  $52 \pm 4$ {\AA}. The estimates of \cite{DimerExp1} and
\cite{Grisenti-exp-2000} make the $^4$He dimer the most extended
known diatomic molecular ground state. The measurements
\cite{Grisenti-exp-2000} also allowed to evaluate a $^4$He-$^4$He
scattering length of $104^{+8}_{-18}$ {\AA} and a $^4$He dimer
energy of $1.1^{+0.3}_{-0.2}$~mK.

%%%%%%%%%%%%%%%   TABLE I: %%%%%%%%%%%%%%%%%%%%%%%%%%%%%%
%\vspace*{-0.5cm}
\begin{table}[h]
\caption{Calculated values of $^4$He dimer energy $\varepsilon_d$,
bond length $\langle R\,\rangle$, and $^4$He atom-atom scattering
length $\ell_{\rm sc}^{(2)}$ for three different atom-atom
potentials, compared to the corresponding experimental values;
estimates for the number $N_\mathrm{Efi}$ of Efimov states based on
formula \eqref{NEfim}.} \label{tableDimerLen}
\begin{center}
\begin{tabular}{ccccc}
\hline \hline
 Potential & $\varepsilon_d$ (mK) &
 $\ell^{(2)}_{\rm sc}$ (\AA) & $\langle R\,\rangle$ (\AA) & ${N_\mathrm{Efi}}^\mathrm{b}$\\[1ex]
\hline
HFD-B  & $-1.68541$ &   $ 88.50$& 46.18 & 0.80\\
LM2M2 & $-1.30348$ &  100.23& 52.00 & 0.83 \\
 TTY   & $-1.30962$ &  100.01& 51.89 & 0.83\\
 \hline
 & & & \\[-2ex]
Experiment$^\mathrm{a}$  & $1.1^{+0.3}_{-0.2}$
 &$104^{+8}_{-18}$ &$52^{+4}_{-4}$\\[1ex]
\hline \hline
\end{tabular}

{\footnotesize $^\mathrm{a}$Reference \cite{Grisenti-exp-2000}.
$^\mathrm{b}$Reference \cite{Aziz95}.}
\end{center}
\end{table}

\begin{table}[htb]
\caption {Ground-state ($E_0$) and excited-state ($E^*$) energies of
the $^4\mathrm{He}$ trimer and the $^{4}$He atom-dimer scattering
length $\ell^{(1+2)}_{\rm sc}$ in case of the TTY atom-atom
potential \cite{TTY}.} \label{tableTrimers}
\begin{center}
\begin{tabular}{cccccccc}
\hline
\hline \\[-2ex]
            & \cite{MSSK} & \cite{Lewerenz} & \cite{BlumeGreene} & \cite{RoudnevYak} &  \cite{Bressanini}
&\  \cite{Barletta}  &\ \cite{SalciLevin}  \\[0.1ex]
\hline \\[-2.5ex]
$\bigl|E_0\bigr|$ (mK) &\  $125.8$ &\ $126.0$  &\ $126.1$ &\ 126.40  &\ 126.4 &\   126.4  & $126.2$\\[1ex]
$\bigl|E^{*}\bigr|$ (mK) &\  $2.282^\mathrm{a}$ &\   &\  &\  2.280  &\    &\  2.277 & \\[1ex]
$\ell^{(1+2)}_{\rm sc}$ (\AA)
&\  $116^\mathrm{b}$ &\   &\  &\  115.8$^\mathrm{c}$   &\  &\  & \\[1ex]
\hline \hline
\end{tabular}

{\footnotesize $^\mathrm{a}$This value was rounded in \cite{MSSK}.
$^\mathrm{b}$Result of extrapolation (see \cite{PRA04}).
$^\mathrm{c}$Result from Ref. \cite{Roudnev1}.}
\end{center}
\end{table}

\begin{table}[htb]
\caption {Ground-state ($E_0$) and excited-state ($E^*$) energies of
the $^4\mathrm{He}$ trimer and the $^{4}$He atom-dimer scattering
length $\ell^{(1+2)}_{\rm sc}$ in case of the LM2M2 atom-atom
potential \cite{Aziz91}.} \label{tableTrimers1}
\begin{center}
\begin{tabular}{cccccccc}
\hline
\hline \\[-2ex]
   & \cite{KolganovaRC2010} &  \cite{RoudnevYak}
   &\cite{Lazauskas} & \cite{MSSK} & \cite{BlumeGreene113}  &
   \cite{Barletta09}    &  \cite{Orlandini2009}  \\[0.1ex]
\hline \\[-2.5ex]
$\bigl|E_0|$ (mK) &  $126.507$ &   $126.41$ &
 126.39 & 125.9 & $126.2$  &   126.15    & $125.6$   \\[1ex]
$\bigl|E^{*}\bigr|$ (mK) &  $2.276$ & 2.271
 & 2.268  & 2.282& 2.26 &  2.274    & $2.245$\\[1ex]
 $\ell^{(1+2)}_{\rm sc}$ (\AA)
&\   &\ $115.4^\mathrm{a}$  &\ 115.56 &\  115$^\mathrm{b}$   &\ 126$^\mathrm{c}$ &\ 120.91 & \\[1ex]
\hline \hline
\end{tabular}
\end{center}
{\footnotesize $^\mathrm{a}$Result from Ref. \cite{Roudnev1}.
$^\mathrm{b}$Result of extrapolation (see \cite{PRA04}).
$^\mathrm{c}$Result from Ref. \cite{BlumeGreene}.}
\end{table}

In 2000, a promising suggestion has been made by Hegerfeldt and
K\"ohler  \cite{Hegerfeldt} concerning the experimental observation
of an Efimov state in $^4$He trimers. The suggestion was to study
diffraction of ultracold $^4$He clusters by inclined diffraction
gratings and look for the specific traces of the excited trimers in
the diffraction picture. The practical realization
\cite{HeEfimovExp} of such an experiment on a grating of a 1000
{\AA} period did not lead to a convincing success. So, a reliable
experimental evidence for the existence of excited states in $^4$He
trimers is still missing. However, in the experiment
\cite{HeEfimovExp} the size of the $^4$He$_3$ ground state has been
estimated for the first time. According to \cite{HeEfimovExp} the
He-He bond length in the $^4$He$_3$ ground state is
$11^{+4}_{-5}$~{\AA}, in agreement with theoretical predictions.

The paper is organized as follows. In Section \ref{S-Efimov} we
recall the basics of the Efimov effect and make historic references
to several approaches that were used to prove this phenomenon,
including the references to rigorous mathematical proofs. Although
this is not directly related to helium trimers, we also give
references to recent experimental works on Efimov physics in
ultracold alcali-atom gases. In Section \ref{S-Ours} we present a
computational evidence for the Efimov nature of the $^3$He trimer
excited state, being based on the investigation in
\cite{KMSa-Brasilia2006,KM-YaF-1999}. Whenever the replacement $V
\to \lambda V$ of a realistic atom-atom potential $V$ is being made,
this consideration shows how the excited state disappears for some
$\lambda>1$. It is absorbed by the continuous spectrum and turned
into a virtual state. For some $\lambda<1$ an additional excited
state pops up, being born from another virtual state. It is this
unusual behavior of the energy levels that indicates that the
$^4$He$_3$ excited state originates due to the Efimov effect.

\section{Efimov effect}
\label{S-Efimov}

The Efimov effect is a remarkable phenomenon that may be viewed as
an excellent illustration for the variety of possibilities arising
when we pass from the two-body to the three-body problem. It is well
known (see, e.g., \cite[Section XIII.3]{ReedSimonIV}) that any
two-particle system with a sufficiently rapidly decreasing and not
too singular interaction $V(\bm{x})$, $\bm{x}\in{\mathbb{R}}^3$, has
a finite number of binding energies. Moreover, the number
$\mathfrak{N}(V)$ of these energies, counting multiplicities,
satisfies the celebrated Birman-Schwinger estimate (see, e.g.,
\cite[Theorem~XIII.10]{ReedSimonIV})
\begin{equation}
\label{BSh} \mathfrak{N}(V)\leq\left(\dfrac{1}{4\pi}\right)^2
\int\limits_{{\mathbb{R}}^3}d\bm{x}\int\limits_{{\mathbb{R}}^3}d\bm{y}
\dfrac{|V(\bm{x})||V(\bm{y})|}{|\bm{x}-\bm{y}|^2}.
\end{equation}
Here, it is assumed that the units are chosen in such a way that the
two-body Schr\"odinger operator in the c.m. system reads
$(H\psi)(\bm{x})=\bigl(-\Delta_{\bm{x}}
+V(\bm{x})\bigr)\psi(\bm{x})$ with $\bm{x}$ the reduced Jacobi
variable and $\Delta_{\bm{x}}$ the Laplacian in $\bm{x}$. Thus, the
convergence of the integral on the r.h.s. part of \eqref{BSh}
ensures that the number of the two-particle binding energies is
finite. In the case of three-particle systems, even with finitely
supported smooth two-body potentials, just the opposite statement
may be true: under certain conditions the number of binding energies
appears to be infinite. Such a spectral situation arises, in
particular, for a system of three spinless particles if none of the
two-body subsystems has bound states but at least two of them have
infinite $s$-wave scattering lengths. This is the essence of the
Efimov effect \cite{VEfimovYaF1970,VEfimov1970}. There is a rigorous
mathematical proof that, for the situation described above, the
number $N(E)$ of three-body binding energies lying below a value
$E<0$ is increasing logarithmically as $E\to0$. Moreover, the
following limit exists \cite{Tamura} (see also \cite{Sobolev})
\begin{equation}
\label{Tamura} \lim\limits_{E\uparrow
0}\dfrac{N(E)}{|\ln|E||}=\Upsilon>0.
\end{equation}
The value of $\Upsilon$ does not depend on details of the (rapidly
decreasing) two-body potentials. It is determined only by the ratios
of particle masses. A qualitative analysis, performed by Efimov
himself in \cite{VEfimovYaF1970,VEfimov1970,VEfimov1973}, allows one
to expect that the following limit exists as well
$$
\lim\limits_{n\to\infty}\dfrac{E_{n+1}}{E_{n}}=\exp\left(-\frac{1}{\Upsilon}\right),
$$
where $E_n$ denotes the bound-state energies numbered in the order
of decreasing absolute values. Furthermore, if the particles are
identical bosons, then Efimov's consideration results in the
following asymptotic relationship
$$
\lim\limits_{n\to\infty}\dfrac{E_{n+1}}{E_n}=\exp(-{2\pi}/{\omega_0})\approx
\frac{1}{515.035},
$$
where  $\omega_0\approx 1.0062378$ is a unique positive solution to
the transcendental equation
\begin{equation}
\label{Enn1} 1-\frac{8}{\sqrt{3}}\, \frac{\mathop{\rm
sinh}\frac{\pi\omega}{6}}{\omega\,\mathop{\rm
cosh}\frac{\pi\omega}{2}}=0.
\end{equation}
This equation first appeared yet in a work \cite{Danilov} by Danilov
on the Skor\-nya\-kov--Ter-Martirosyan equation \cite{STM}. Some
rigorous statements on a more detailed asymptotic behavior of the
Efimov energy levels $E_n$, $n\to\infty$, in a system of three
identical bosons  can be found in \cite{AlbLakMak}. The analysis of
\cite{AlbLakMak} follows an alternative approach to justify the
Efimov effect, the one that was proposed independently by Faddeev in
\cite{FaddeevMIFI} and Amado and Noble in
\cite{AmadoNoble1971,AmadoNoble1972} soon after the papers
\cite{VEfimovYaF1970,VEfimov1970} were published. This approach
involves an explicit separation of a non-Fredholm (as
$\ell^{(2)}_\mathrm{sc}\to\infty$) component of the integral
operator entering the momentum-space Faddeev equations and the
subsequent examination of the three-body spectrum generated by that
component (see \cite[pp. 103--105]{MF}). Notice that the first
completely rigorous proof of the existence of the Efimov effect,
given by Yafaev in \cite{Yafaev}, also follows the approach of
\cite{FaddeevMIFI,AmadoNoble1971,AmadoNoble1972}. For rigorous
results on Efimov properties of $N$-body systems with $N\geq 4$ we
refer to \cite{Zhislin1982,Wang2004,Wang2005}.

All known two-body systems (both nuclear and atomic) have finite
scattering lengths. Therefore, in general it is impossible to
observe the genuine ``full-scale'' Efimov effect (with an infinite
number of three-body bound states). Nevertheless, systems featuring
at least some peculiarities of the Efimov effect are also of great
interest. A qualitative analysis, for the system of three identical
bosons, performed by Efimov in \cite{VEfimovYaF1970,VEfimov1970}
(see also the review paper by Fillips \cite{Phillips1977}) shows
that if the boson-boson scattering length $\ell_{\rm sc}^{(2)}$ is
large compared to the effective radius $r_0$ of the two-body forces,
then there is an effective
 $1/\rho^2$-type attractive interaction on a scale of
$r_0\lesssim\rho\lesssim \ell_{\rm sc}^{(2)}$ where $\rho$ is the
system hyperradius. This conclusion is used as an argument to
approve the following estimate (see \cite{VEfimovYaF1970}):
\begin{equation}
\label{NEfim}
N_\mathrm{Efi}\simeq\dfrac{\omega_0}{\pi}\ln\left|\dfrac{\ell_{\rm
sc}^{(2)}}{r_0}\right|,
\end{equation}
where $N_\mathrm{Efi}$ denotes the total number of bound states in
the three-boson system under consideration. Surely, this estimate is
assumed to work only for very large ratios $\ell_{\rm
sc}^{(2)}/{r_0}$ but it may provide a hint also for relatively small
$\ell_{\rm sc}^{(2)}/{r_0}$. Based on \eqref{NEfim}, for the $^4$He
three-atomic system with realistic atom-atom potentials one
typically obtains $N_\mathrm{Efi}\simeq 0.6$---$1.3 $ (see
\cite{Aziz95}; cf. Table \ref{tableDimerLen}). Since this value is
only around (or even less than) unity, it neither supports nor
disproves the claim that the excited state of the $^4$He trimer is a
genuine Efimov state, and
a further investigation is needed (see Section \ref{S-Ours}).  %%%\div

We also notice that, if the two-body scattering lengths are
infinite, introduction of three-body forces (provided they are short
range) does not affect the Efimov effect, because of its long-range
nature, and the number of binding energies remains infinite. But if
the Efimov effect is not full-scale, i.e. the two-body scattering
lengths are large but finite, the appropriately chosen positive
definite three-body interation may, of course, completely eliminate
any binding in the three-body system.

Already equation \eqref{Enn1} drops a hint that there should be a
close link between the Efimov effect and the Thomas effect
\cite{Thomas}. Recall that the origin of the Thomas effect lies in
the fact that, in the case where two-body interactions are
zero-range, the three-boson Shr\"odinger operator is not
semi-bounded from below \cite{MinlosFaddeev1961}. Hence, there is a
possibility of a collapse of the system with all three particles
falling to the center of mass. Virtually, the asymptotic estimate
\eqref{NEfim} explains both the effects at once. For $r_0\neq 0$ and
$|\ell_{\rm sc}^{(2)}|\to\infty$ it gives us the number of Efimov
levels, which accumulate exponentially towards the three-body
breakup threshold. On the other hand, if $\ell_{\rm sc}^{(2)}$ is
finite and nonzero, this estimate describes the number of energy
levels in the Thomas effect going to $-\infty$ as $r_0\to 0$. That
the Efimov and Thomas effects are nothing but the two sides of the
same coin was noted, in particular, in \cite{AlbHWu} and
\cite{MakarovMelezhik} (see also \cite[Section 5]{NielsenRev2001}
and references therein). Currently, there are a lot of discussions
on the universal properties of three-body systems at ultralow
energies, and there is a tendency (see, e.g.
\cite{Yama2002,LeeKoehler2007}) to use a joint term ``Thomas-Efimov
levels'' for the discrete spectrum arising in both effects. Various
three-body universality aspects and the Efimov effect itself are
discussed in great detail in the advanced review article by Braaten
and Hammer \cite{BraHam2006}.

Till now we only talked on isolated three-particle systems that do
not interact with the rest of the world. It is usually assumed,
explicitly or implicitly (see, e.g. \cite{Beda,Stoll2005,Jonsell,Kraemer}),
that the estimates like \eqref{NEfim} are also valid for three-atom
systems put into an external magnetic field. It is known that, being
subject to a magnetic field, certain two-atom systems experience a
Feshbach resonance due to Zeeman interaction \cite{Moerdijk}. In
such a case one gets an opportunity to control the atom-atom
scattering length, by changing the intensity of the magnetic field.
This is particularly relevant for systems composed of alkaline
atoms. In 2006, the results of an experiment on three-body
recombination in an ultracold gas of cesium atoms have been
announced \cite{Kraemer,Naegerl}. Those results were interpreted by
the authors of the experiment as evidence of the emergence of at
least one Efimov state in the $^{133}$Cs three-atomic system as the
magnetic field appropriately changes. A discussion and different
interpretations of the experiment \cite{Kraemer,Naegerl} can be
found in \cite{LeeKoehler2007,EsryGreene2006N}. An experimental
evidence for the Efimov resonant states in heteronuclear three-atom
systems consisting of $^{41}K$ and $^{87}$Rb atoms was reported in
\cite{Barontini2009}. Recently, signatures of the Efimov effect have
been found experimentally in a three-component gas consisting of
$^6$Li atoms that are settled in the three different lowest-energy
states~\cite{Lompe2010}.

\section{On the Efimov nature of the $^4$He trimer excited state}
\label{S-Ours}

Although quite different Aziz \textit{et al.} atom-atom potentials
\cite{Aziz79,Aziz87,Aziz91} and very different numerical techniques
were used in \cite{KMS-JPB,Gloeckle,EsryLinGreene}, the main
conclusions concerning the trimer excited state are basically the
same. Namely, this state disappears if the potential is multiplied
by a factor $\lambda$ of about 1.2.  More precisely, if the
atom-atom potential is multiplied by $\lambda>1$ then the following
effect is observed. First, with increasing $\lambda$ the trimer
excited state energy $E^*(\lambda)$ goes deeper more rapidly than
the dimer energy $\varepsilon_d(\lambda)$, i.e. the difference
$\varepsilon_d(\lambda)-E^*(\lambda)$ increases. At some point the
behavior of this difference changes to the opposite, that is, with
further increase of $\lambda$ it decreases monotonously. In other
words, from now on the dimer energy $\varepsilon_d(\lambda)$ goes
down quicker than the ex\-ci\-ted-sta\-te energy $E^*(\lambda)$. At
$\lambda\approx 1.2$ the level $E^*$ disappears, being covered by
the continuous spectrum. It is just such a nonstandard behavior of
the energy $E^*(\lambda)$ that points to the Efimov nature of the
trimer excited state. Vice versa, if $\lambda$ slightly decreases
from 1 by not more than 2\,\%, the second excited state $E^{**}$
shows up \cite{Gloeckle,EsryLinGreene}. In \cite{Lazauskas} and
\cite{KMS-FBS2008} the Efimov nature of the $^4$He trimer excited
state was discussed in terms of the atom-atom scattering length.

Apparently, the most detailed numerical study of the nature of the
excited state in the $^4$He trimer has been performed in
\cite{KM-YaF-1999} (see also \cite{KM-CPC-2000} and
\cite{MoK-FBS-1999}). Notice that the Aziz \textit{et al.} potential
\cite{Aziz87} was employed in \cite{KM-YaF-1999} and the number of
the partial-wave Faddeev components was reduced to one. This,
however, should not affect the basic qualitative conclusions. One of
the goals of \cite{KM-YaF-1999} was to elucidate the fate of the
excited state, once it leaves the physical sheet (at some
$\lambda>1$). Another goal was to study the emergence mechanism for
new excited states as $\lambda$ ($\lambda<1$) is decreasing.

It was found in \cite{KM-YaF-1999} that, for $\lambda$ ($\lambda>1$)
increasing, the trimer excited-state energy $E^*(\lambda)$ merges
with the two-body threshold $\varepsilon_d(\lambda)$ at
$\lambda\approx1.175$. As the factor $\lambda$ decreases further, it
transforms into a first-order virtual level. New excited-state
energy levels at $\lambda<1$ emerge from the first-order virtual
levels as well. The latter show up in pairs. The emergence of a pair
of first-order virtual levels is preceded by a collision and
subsequent fusion of a pair of conjugate first-order resonances into
a second-order virtual level. It is worth to notice, however, that
these resonances may not be the true resonances, since they are
lying outside the energy region where the applicability of the
computational approach of \cite{KM-YaF-1999} was proven to work (see
\cite{Mot-MNach1997}).

\begin{table}
[htb] \caption{$^4$He dimer binding energy $\varepsilon_d$, energies
of the first ($E^*$) and second ($E^{**}$) excited states of the
$^4$He trimer; virtual-state energy $E_{\rm virt}$ of the $^4$He
three-atom system; $^4$He atom-atom and $^4$He atom-dimer scattering
lengths $\ell^{(2)}_{\rm sc}$  and $\ell_{\rm sc}^{(1+2)}$,
respectively, as functions of the potential strength factor
$\lambda$. All energies are given in mK, the scattering lengths in
{\AA}. The dashes mean the nonexistence of the corresponding states.
The HFD-B atom-atom potential \cite{Aziz87} was used in the
computations. } \label{tableScLength1}

\begin{center}
\begin{tabular}{|l|ccccrc|}
\hline\hline $\qquad\lambda$ & $\varepsilon_d$ & $\varepsilon_d -
E^*$ & $\varepsilon_d - E_{\rm virt}$ & $\varepsilon_d - E^{**}$ &
$\ell_{\rm sc}^{(1+2)}$ & $\ell^{(2)}_{\rm sc}$   \\
\hline
\quad 1.30 & $-199.45$ & - & 1.831 & - & $-61$ & 11.4  \\
\quad 1.20 & $-99.068$ & - & 0.01552 & - & $-340$ & 14.7  \\
\quad 1.18 & $-82.927$ & - & 0.00058 & - & $-1783$ & 15.8  \\
\quad 1.17 & $-75.367$ & 0.0063 & - & - & 8502 & 16.3  \\
\quad 1.15 & $-61.280$ & 0.0737 & - & - & 256 & 17.7  \\
\quad 1.10 & $-32.222$ & 0.4499 & - & - & 152 & 23.1  \\
\quad 1.0 & $-1.685$ & 0.773 & - & - & 160 & 88.6  \\
\quad 0.995 & $-1.160$ & 0.710 & - & - & 151 & 106  \\
\quad 0.990 & $-0.732$ & 0.622 & - & - & 143 & 132 \\
\quad 0.9875$\quad$ & $-0.555$ & 0.573 & 0.222 & - & 125 & 151  \\
\quad 0.985 & $-0.402$ & 0.518 & 0.097 & - & 69 & 177  \\
\quad 0.982 & $-0.251$ & 0.447 & 0.022 & - & $-75$ & 223  \\
\quad 0.980 & $-0.170$ & 0.396 & 0.009 & - & $-337$ & 271 \\
\quad 0.9775 & $-0.091$ & 0.328 & 0.003& - & $-6972$ & 370 \\
\quad 0.975 & $-0.036$ & 0.259 & - & 0.002 & 7120 & 583 \\
\quad 0.973 & $-0.010$ & 0.204 & - & 0.006 & 4260 &  1092  \\
\hline\hline
\end{tabular}
\end{center}
\end{table}

As an illustration of what has been said above, we present Table
\ref{tableScLength1} taken from \cite{KMSa-Brasilia2006}. It is seen
that for $0.9875<\lambda\leq 1.17$ the $^4$He trimer has only one
excited state of energy $E^*$ (see the third column). For
$\lambda\geq 1.18$, instead of the excited state a virtual state of
energy $E_{\rm virt}$ shows up (see the fourth column). This occurs
as a consequence of the excited-state energy passing to the
unphysical sheet.

As $\lambda$ decreases down to approximately 0.986, a new virtual
level arises (see the fourth column again). We use the same notation
$E_{\rm virt}$ for the energy of that level. A further decrease of
the factor $\lambda$ to approximately 0.976 shifts the virtual level
$E_{\rm virt}$ to the physical sheet, which results in the emergence
of the second excited state (see the fifth column). The binding
energy of this state is denoted by $E^{**}$.

\begin{figure}
%[htb]
\begin{center}
{\includegraphics[angle=0,width=12.8cm]{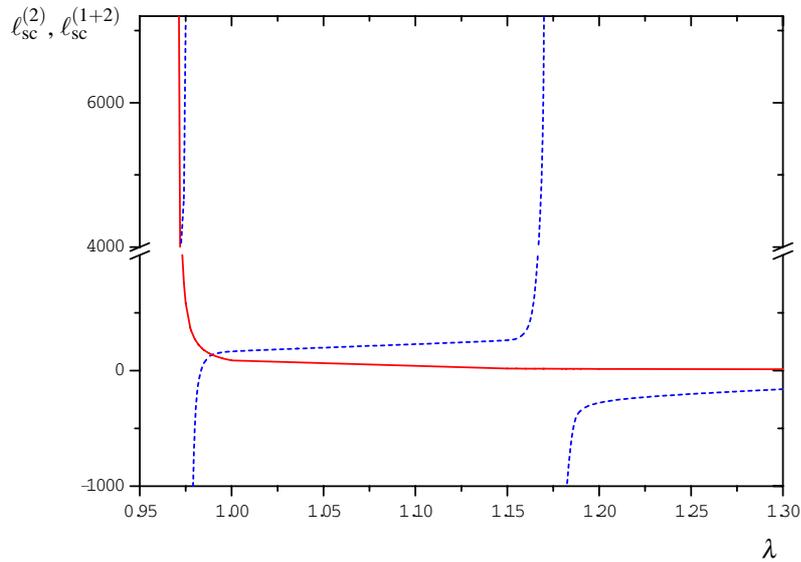}}
\end{center}
\vspace*{-8.3cm} \hspace*{6.em} $\ell^{(2)}_{\mathrm{sc}}$,
$\ell^{(1+2)}_{\mathrm{sc}}$ \vspace*{6.6cm}

\hspace*{12.0cm}$\lambda$

\caption{Dependence of the $^4$He atom-atom scattering length
$\ell_{\mathrm{sc}}^{(2)}$ (\AA) (solid curve) and the $^4$He
atom-dimer scattering length $\ell_{\mathrm{sc}}^{(1+2)}$ (\AA)
(dashed curve) on the potential strength factor~$\lambda$. The HFD-B
atom-atom potential of Ref. \cite{Aziz87} was used in the
computations.} \label{ell-lambda}
\end{figure}

In both of the above cases, the transformation of a virtual state
into an excited state changes the sign of the atom-dimer scattering
length $\ell^{(1+2)}_{\rm sc}$. At the corresponding values of
$\lambda$ the function $\ell^{(1+2)}_{\rm sc}(\lambda)$ has
pole-like singularities (see the sixth column of Table
\ref{tableScLength1}) while the atom-atom scattering length
$\ell^{(2)}_{\rm sc}$ varies continuously and monotonously. The
behavior of both the scattering lengths $\ell^{(2)}_{\rm
sc}(\lambda)$ and $\ell^{(1+2)}_{\rm sc}$ shown in Table
\ref{tableScLength1} is graphically displayed in Fig.
\ref{ell-lambda}.

\section{Conclusion}
\label{S-Concl}

We have reviewed results obtained in the last forty years which
prove the Efimov nature of the $^4$He three-atomic system. This
system appears to be the best, most thoroughly investigated example
where the Efimov effect manifests itself. According to what is
shown, the most vital questions in this context have been asked and
answered. There are, of course, numerous other questions that
concern, e.g., the Efimov aspects of larger He$_n$ systems, the
influence of external fields, the properties of mixed atomic systems
\text{etc.} All this is the topic of further investigations based on
Efimov's fundamental idea (see, e.g.,
\cite{Stecher2010,Esry2009,WangEsry2009,Ferlaino2010} and references therein).

%%\footnotesize

%%\newpage

\end{document}